\documentclass[onecolumn,superscriptaddress,pra]{revtex4}%

  \expandafter\let\csname equation*\endcsname\relax 
  \expandafter\let\csname endequation*\endcsname\relax

\usepackage{theorem}
\usepackage{enumerate}
\usepackage{amsmath}
\usepackage{amssymb}
\usepackage{graphicx}
\usepackage{epstopdf}
\usepackage{color}
\usepackage{euscript}
\usepackage{float}

\theorembodyfont{\rmfamily}
\newtheorem{Theorem}{Theorem}
\newtheorem{Proposition}{Proposition}
\newtheorem{Definition}{Definition} 

\newtheorem{Lemma}{Lemma}
\newtheorem{Proof}{Proof}
\newtheorem{Conjecture}{Conjecture}

\newcommand{\lv}{\left \vert}
\newcommand{\rv}{\right \vert}

\newcommand{\ra}{\right \rangle}
\newcommand{\ket}[1]{\lv #1 \ra}

\newcommand{\braket}[2]{\langle #1 \vert #2 \rangle}
\newcommand{\ketbra}[2]{\lv #1 \rangle \langle #2 \rv}

\newcommand{\tr}{\mathrm{tr}}

\begin{document}

\title{Generating a state $t$-design by diagonal quantum circuits}

\author{Yoshifumi Nakata}

\address{
Institute for Theoretical Physics, Leibniz University Hannover, Germany\\
Department of Physics, Graduate School of Science,
University of Tokyo, Tokyo, Japan}
\email{yoshifumi.nakata@itp.uni-hannover.de}

\author{Masato Koashi}
\address{Photon Science Center, University of Tokyo, Tokyo, Japan}

\author{Mio Murao}

\address{Department of Physics, Graduate School of Science,
University of Tokyo, Tokyo, Japan\\
Institute for Nano Quantum Information Electronics,
 University of Tokyo, Tokyo, Japan
}

\begin{abstract}
We investigate protocols for generating a state $t$-design by using a fixed separable initial state and 
a diagonal-unitary $t$-design in the computational basis, which is a $t$-design
of an ensemble of diagonal unitary matrices with random phases as their eigenvalues.
We first show that a diagonal-unitary $t$-design generates a $O(1/2^N)$-approximate state $t$-design, where $N$ is the number of qubits. 
We then discuss a way of improving the degree of approximation by exploiting non-diagonal gates after applying a diagonal-unitary $t$-design.
We also show that it is necessary and sufficient to use $O(\log_2 t)$-qubit gates with random phases to generate a diagonal-unitary $t$-design by diagonal quantum circuits,
and that each multi-qubit diagonal gate can be replaced by a sequence of multi-qubit controlled-phase-type gates with discrete-valued random phases.
Finally, 
we analyze the number of gates for implementing a diagonal-unitary $t$-design by {\it non-diagonal} two- and one-qubit gates.
Our results provide a concrete application of diagonal quantum circuits in quantum informational tasks.
\end{abstract}

\maketitle

\section{Introduction}

Diagonal quantum circuits in the computational basis are recently attracting much attention~\cite{SB2009,BJS2011,NV2012,JWAURB2013,FM2013}.
Despite the commutativity of diagonal gates,
it has been shown that they are likely to have stronger computational power than classical computers even if the initial state is fixed to be a separable state.
This implies that the commutativity of gates does not immediately result in a trivial computational power.
From an experimental point of view, diagonal gates can be fault-tolerantly realized in, e.g., super- and semi-conducting systems~\cite{ABDPST2009}.
Hence, any applications of diagonal circuits will lead to an experimental demonstration of a quantum advantage in informational tasks,
but little is known about concrete applications of diagonal circuits so far.

One of the applications of diagonal circuits, proposed by two of the authors~\cite{NM2012},
is related to random states, which are an ensemble of pure states uniformly distributed in a Hilbert space with respect to the unitarily invariant measure.
They have many utilities in a wide range of applications, e.g.,
in quantum communicational tasks~\cite{L1997}, for efficient measurements~\cite{RBSC2004}, for an algorithmic use~\cite{RRS2005,S2006} and for estimation of gate fidelities~\cite{DCEL2009}.
Despite such applications, exact random states cannot be efficiently generated.
Hence, efficient generations of a $t$-design of random states, called a {\it state} $t$-design, using quantum circuits have been intensely studied~\cite{DCEL2009,EWSLC2003,DLT2002,ODP2007,DOP2007,Z2008,HL2009,HL2009TPE,BH2010,BHH2012,CHMPS2013},
where a $t$-design of an ensemble is an ensemble that simulates up to $t$th-order statistical moments of the original ensemble~\cite{RBSC2004,DCEL2009,AE2007}.
In most applications of random states, a state $t$-design for small $t$ is sufficient~\cite{L2009} 
and it has been shown that an approximate state $t$-design can be efficiently generated by a quantum circuit called a {\it local random circuit}~\cite{BHH2012}.
In Ref.~\cite{NM2012}, a protocol has been proposed for generating an exact state $2$-design
by combining a diagonal quantum circuit with a classical probabilistic procedure,
which provides a usage of diagonal quantum circuits that leads to several applications in quantum tasks.

In this paper, we investigate 
protocols of generating a state $t$-design for general $t$
by using a $t$-design of random diagonal-unitary matrices called a {\it diagonal-unitary $t$-design}.
We first show that a good approximate state $t$-design is obtained simply by applying a diagonal-unitary $t$-design in the computational basis
to a fixed separable state. 
The degree of approximation is given by $t(t-1)/2^N + O(1/2^{2N})$ for a constant $t$, where $N$ is the number of qubits.
This result is interesting from two perspectives.
From a theoretical point of view, it shows that a diagonal quantum circuit 
can generate an ensemble of states whose distribution in a Hilbert space is hard to be distinguished from the uniform one as long as looking at 
lower order statistical moments.
This may help an intuitive understanding of a strong computational power of diagonal quantum circuits. 
On the other hand, from experimental point of view, our protocol extends a usage of diagonal quantum circuits in quantum applications and 
can be used for demonstrating a quantum advantage.

We also study a way of improving the degree of approximation
by using a local random circuit after applying a diagonal-unitary $t$-design.
Since an ensemble of states after a diagonal-unitary $t$-design is already a good approximate state $t$-design,
it is natural to expect that this protocol has an advantage to reduce the number of gates in the local random circuit
compared to
the one that uses only a local random circuit. We numerically confirm that this seems to be the case.

It is also important to investigate efficient implementations of a diagonal-unitary $t$-design by quantum circuits.
Although a diagonal-unitary $t$-design contains only diagonal matrices, 
it cannot be implemented in general by using only two- and one-qubit diagonal gates
since multi-qubit diagonal gates are generally not decomposable into diagonal gates acting on smaller number of qubits.
For instance, a three-qubit gate ${\rm diag} (1,1,1,1,1,1,1,-1 )$ cannot be represented by a quantum circuit consisting of only two- and one-qubit diagonal gates.   This indecomposability of multi-qubit diagonal gates into two-qubit diagonal gates should be contrasted with the decomposability of multi-qubit general unitary gates into two- and one-qubit unitary gates.
We show that, if we use only diagonal gates, it is necessary and sufficient to use $(\lfloor \log_2 t \rfloor+1)$-qubit gates with random phases
for generating a diagonal-unitary $t$-design, where $\lfloor x \rfloor$ denotes the maximum integer that does not exceed $x$.
We also show that the multi-qubit diagonal gates in the circuit for implementing a diagonal-unitary $t$-design can be replaced 
by multi-qubit {\it controlled-phase-type} gates with {\it discrete} random phases.
We finally discuss how to generate a diagonal-unitary $t$-design by using {\it non-diagonal} two-qubit gates, and provide a construction of a quantum circuit 
implementing a diagonal-unitary $t$-design by $O(N^{\log_2 t})$ two-qubit gates for a constant $t$, while it will not be optimal.

Before leaving the introduction, we would like to note that partially randomizing unitary matrices while preserving some properties,
which is the case in a diagonal-unitary $t$-design, 
is not necessarily simpler than full randomization in the unitary group, which is a concern of a unitary $t$-design. 
For a partial randomization, we need to perform two conflict tasks, {\it randomization} and {\it preservation}, at the same time.
Thus, although commutativity of diagonal gates simplifies an investigation of random diagonal-unitary matrices,
it is not trivial whether an implementation of diagonal-unitary $t$-design is simpler than that of a unitary $t$-design.

This paper is organized as follows.
In Sec.~\ref{Sec:RSPRS}, we review the definitions of terms used in this paper.
We summarize all of our main results in Sec.~\ref{S:MR}. Their proofs are provided in Sec.~\ref{S:DandP}.
We make concluding remarks in Sec.~\ref{S:Conc}.

\section{Random unitary matrices and $t$-designs} \label{Sec:RSPRS}

We first review the definitions of random unitary and diagonal-unitary matrices~\cite{M1990,NM2012}, random and phase-random states~\cite{NTM2012}, and their $t$-designs~\cite{AE2007}.
In the following, we denote by $\ket{0}$ and $\ket{1}$ the computational basis of the Hilbert space of a qubit, which are the eigenstates of the Pauli $Z$ operator with eigenvalues
$+1$ and $-1$, respectively.
For simplicity, we also denote by $\mathbb{E}$ expectations over a probability distribution. If necessary, we specify the probability space taken over for the expectation.

\begin{Definition}[Random unitary matrices and random states]
Let $\mathcal{U}(d)$ be the unitary group of degree $d$.
{\it Random unitary matrices} $\mathcal{U}_{\rm Haar}$ are the ensemble of unitary matrices uniformly distributed with respect to the Haar measure on $\mathcal{U}(d)$. {\it Random states} $\Upsilon_{\rm Haar}$ are the ensemble of states $\{ U \ket{\Psi} \}_{U \in \mathcal{U}_{\rm Haar}}$ for any fixed state $\ket{\Psi} \in \mathcal{H}$,
where $\mathcal{H}$ is a Hilbert space with dimension $d$.
\end{Definition}

\begin{Definition}[Random diagonal-unitary matrices and phase-random states]
{\it Random diagonal-unitary matrices} in an orthonormal basis $\{ \ket{u_n} \}$ denoted by $\mathcal{U}_{\rm diag}(\{ \ket{u_n} \})$ are an ensemble of diagonal unitary matrices of the form $U_{\varphi} =  \sum_{n=1}^{d} e^{i \varphi_n} \ketbra{u_n}{u_n}$, where the phases $\varphi_n$ are uniformly distributed according to the normalized Lebesgue measure d$\varphi$ = d$\varphi_1 \cdots $d$\varphi_{d} / (2\pi)^{d} $ on $[0,2\pi)^d$.
{\it Phase-random states} $\Upsilon_{\rm phase}(\{ | \braket{u_n}{\Psi} |, \ket{u_n } \})$ are an ensemble of states $\{ U \ket{\Psi} \}_{U \in \mathcal{U}_{\rm diag}(\{ \ket{u_n} \})}$.
\end{Definition}

Note that a distribution of random states is independent of an initial state $\ket{\Psi}$ due to the unitary invariance of the Haar measure.
This is not the case in phase-random states and their distribution depends on the initial state.

A $t$-design of an ensemble is defined by an ensemble that simulates up to the $t$th-order statistical moments of the original ensemble on average~\cite{RBSC2004,DCEL2009,AE2007,TG2007}.
Although a $t$-design is required to be a finite ensemble in several definitions, we do not require it to be more general.

\begin{Definition}[$\epsilon$-approximate unitary $t$-designs]
Let $\mathcal{U}$ be random unitary matrices or random diagonal-unitary matrices.
An $\epsilon$-approximate $t$-design of $\mathcal{U}$, denoted by $\mathcal{U}^{(t,\epsilon)}$, is an ensemble of unitary matrices such that
\begin{equation*}
\biggl|\!\biggl| 
\mathbb{E}_{U \in \mathcal{U}^{(t,\epsilon)}} [ U^{\otimes t} \otimes (U^{\dagger})^{\otimes t}]
 - 
\mathbb{E}_{U \in \mathcal{U}} [ U^{\otimes t} \otimes (U^{\dagger})^{\otimes t}]
 \biggr|\! \biggr|_1 \leq \epsilon,
\end{equation*}
where $|\!| \cdot |\!|_{1} = \tr | \cdot |$ is the trace norm.
The $t$-designs for random unitary and diagonal-unitary matrices are called {\it unitary} and {\it diagonal-unitary} $t$-designs, respectively. 
\end{Definition}

\begin{Definition}[$\epsilon$-approximate state $t$-designs] \label{Def:appstate}
Let $\Upsilon$ be random states or phase-random states.
An $\epsilon$-approximate $t$-design of $\Upsilon$, denoted by $\Upsilon^{(t,\epsilon)}$, is an ensemble of states such that
\begin{equation*}
\biggl|\!\biggl|
\mathbb{E}_{\ket{\psi} \in \Upsilon^{(t,\epsilon)}} [ \ketbra{\psi}{\psi}^{\otimes t} ] 
- 
\mathbb{E}_{\ket{\psi} \in \Upsilon} [\ketbra{\psi}{\psi}^{\otimes t}]
 \biggr|\! \biggr|_{1} \leq \epsilon.
\end{equation*}
In particular, we call a $t$-design of random states a state $t$-design in this paper.
\end{Definition}

A $t$-design for $\epsilon=0$ is called an {\it exact} $t$-design.
We mean the exact ones when we simply use the term $t$-designs in this paper.
Although we have presented definitions of $\epsilon$-approximate $t$-designs in terms of the trace norm, there are other definitions using different distance measures such as the diamond norm
and the Schatten norms (see, e.g., Ref.~\cite{L2010}).
However, they are shown to be all equivalent,namely, if $\mathcal{V}$ is an $\epsilon$-approximate $t$-design in one of the definitions,
then it is also an $\epsilon'$-approximate $t$-design in other definitions, where $\epsilon' = {\rm poly} (2^{tN}) \epsilon$~\cite{L2010}.

A unitary and a state $t$-design can be used in many quantum informational tasks.
For instance, random states saturate the classical communication capacity of a noisy quantum channel~\cite{L1997},
and are also related to optimal measurements in tomography~\cite{RBSC2004}. 
POVM measurements in a random basis can be used for solving hidden subgroup problems efficiently~\cite{RRS2005}.
Random states in these use can be replaced by a state $t$-design for a small $t$~\cite{L2009}.
A $2$-design of random states is known to be useful for checking the fidelity of quantum gates~\cite{DCEL2009}.

\section{Main results} \label{S:MR}

We summarize our results in this section. 
In Subsec.~\ref{SS:App}, 
we provide protocols of generating a state $t$-design by using a diagonal-unitary $t$-design in the computational basis.
Our results about implementations of diagonal-unitary $t$-designs by diagonal circuits are presented in Subsec.~\ref{SS:Imp}.  All the proofs are given in Section 4.

\subsection{Protocols of generating a state $t$-design by using a diagonal-unitary $t$-design} \label{SS:App}

Applying a diagonal-unitary $t$-design on any pure state achieves a $t$-design of the corresponding phase-random states by definition. 
If we choose an appropriate initial state and a basis of the diagonal-unitary $t$-design,
we can also achieve a good approximation of a $t$-design of {\it random states}, as stated in the following Proposition:

\begin{Proposition} \label{Prop:Protocol1}
A $t$-design of phase-random states obtained
by applying a diagonal-unitary $t$-design in the computational basis onto an initial state $\ket{+}^{\otimes N}$, where $\ket{+}=\frac{1}{\sqrt{2}}(\ket{0}+\ket{1})$,
is an $\eta(N,t)$-approximate state $t$-design, where $\eta(N,t)=\frac{t(t-1)}{d} + O(\frac{1}{d^2})$ and $d=2^N$.
\end{Proposition}

As we will see in the next subsection, a diagonal-unitary $t$-design for a small $t$ can be achieved by using only diagonal gates
acting on a small number of qubits,
where the order of the applications of gates does not matter due to the commutativity of diagonal gates, i.e., there is no
inherent temporal structure in the circuit. This is a big advantage in experimental implementations of the circuit.
In particular, a diagonal-unitary $t$-design for $t \leq 3$ is implementable by using two- and one-qubit diagonal gates and the total number of gates is $O(N^2)$.
This means that the protocol in Proposition~\ref{Prop:Protocol1} generates an $\eta(N,t)$-approximate state $t$-design for $t \leq 3$ 
by a quantum circuit composed of $O(N^2)$ two- and one-qubit diagonal gates that has no temporal structure.
This should be contrasted to a previously known protocol using a {\it local random circuit}~\cite{BHH2012},
which is composed of two-qubit gates randomly chosen from $U(4)$ acting only on neighboring qubits.
Although it achieves the same degree of approximation by using at most $O(t^5 \log(t) N^2)$ gates,
the circuit is necessarily temporally structured.
Thus, our protocol has a practical advantage for large $N$ as long as the required degree of approximation is up to $\eta(N,t)$,
particularly when $t\leq 3$.

To achieve a better degree of approximation, it may help to combine a diagonal-unitary $t$-design with other procedures.
In Ref.~\cite{NM2012}, it was shown that an exact state $2$-design is obtained if we 
combine a diagonal-unitary $2$-design with a classical probabilistic procedure.
However, we can show that adding a classical probabilistic procedure improves the degree of approximation by $O(d^{1-t})$ in general,
so that it is not effective for $t \geq 3$ (see~\ref{Ap:classical}).
Here, we examine a method to apply a local random circuit following the application of
a diagonal-unitary $t$-design on $\ket{+}^{\otimes N}$. 
Since an $\eta(N,t)$-approximate state $t$-design is already achieved by a diagonal-unitary $t$-design,
it is natural to expect that this protocol generates an $\epsilon$-approximate state $t$-design more efficiently than the one that uses only a local random circuit.
We numerically check this.
In contrast to the previous results about a local random circuit, where an input state is arbitrary,
input states of the local random circuit in our protocol are determined by output states obtained by applying a diagonal-unitary $t$-design on $\ket{+}^{\otimes N}$.
Hence, the necessary length of the local random circuit in our protocol is not directly obtained from the previous results.

\begin{figure}[tb!]
\centering
\includegraphics[width=75mm, clip]{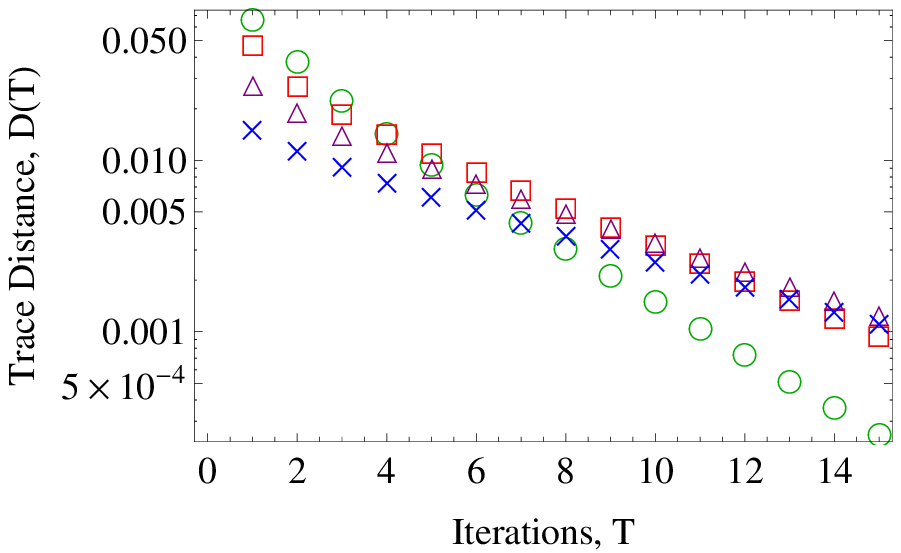} 
\includegraphics[width=75mm, clip]{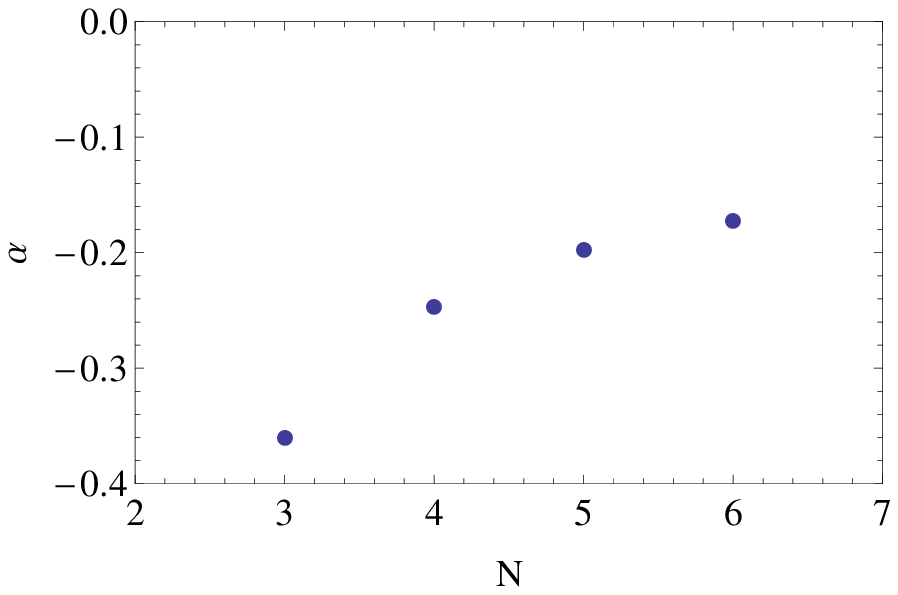} 
  \caption{ The left figure shows how $D(T)$ changes when we apply a parallelized local random circuit of length $T$ 
after the application of a diagonal-unitary $t$-design on the initial state $\ket{+}^{\otimes N}$. 
The numerics is performed by Mathematica 9 for $t=2$ and each plot $\circ$ (green), $\Box$ (red), $\triangle$ (purple), and $\times$ (blue), represents $N=3,4,5,6$, respectively. 
The number of sampling at each point of the plot is $1000$. 
The right figure shows how the coefficient $\alpha$ depends on $N$.}
\label{Fig:TD}
\end{figure}

Let $T$ be the length of a parallelized local random circuit after applying a diagonal-unitary $t$-design,
where we mean by a parallelized local random circuit that unitary gates acting on different qubits are applied simultaneously.
We denote by $D(T)$ the trace distance between an expectation of $\ketbra{\phi}{\phi}^{\otimes t}$ over the resulting ensemble and that over a state $t$-design
(see Definition~\ref{Def:appstate}).
We numerically check how $D(T)$ scales with $T$ for $t=2$ and $N=3,4,5,6$.
In the numerics, we randomly generate a unitary matrix representing a parallelized local random circuit, and apply it to the states obtained 
by applying a diagonal-unitary $t$-design on $\ket{+}^{\otimes N}$. By repeating this and averaging the resulting states, 
we evaluate the expectation of $\ketbra{\phi}{\phi}^{\otimes t}$ over the ensemble obtained by our protocol.
The result is shown in Fig.~\ref{Fig:TD}.
We observe that the distance exponentially decreases for each $N$. 
Although the exponent of the exponential decrement depends on $N$ for up to $N=6$, we expect from the right figure in Fig.~\ref{Fig:TD} that it converges to some value $
\alpha(t)$ that depends on $t$ but not on $N$.
Accordingly, we conjecture the following:
\begin{Conjecture}
Let $D(T)$ be the trace distance as defined above. Then,  
\begin{equation*}
D(T) \sim \eta(N, t) 2^{ - T/ \alpha(t)} \sim \frac{t(t-1)}{d} 2^{- T/\alpha(t)}.
\end{equation*}
\end{Conjecture}

If this is the case, a parallelized random circuit of length $T(\epsilon)$ following a diagonal-unitary $t$-design 
achieves an $\epsilon$-approximate state $t$-design, where
\begin{equation*}
T(\epsilon) = \begin{cases}
0 & \text{ for } \epsilon \geq \eta(N,t), \\
\alpha(t) [\log_2 1/\epsilon - N + \log_2 t(t-1)] & \text{ for } \epsilon < \eta(N,t).
\end{cases}
\end{equation*}

\subsection{Implementations of diagonal-unitary $t$-designs} \label{SS:Imp}

\begin{figure}[tb!]
\centering
\includegraphics[width=75mm, clip]{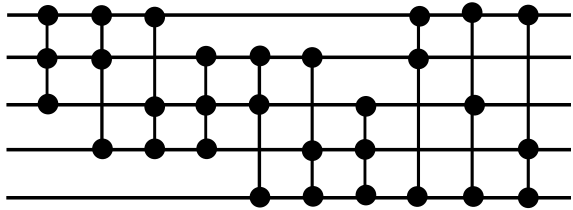} 
  \caption{ An $r$-qubit phase-random circuit for an $N$-qubit system where $N=5$ and $r=3$. The black circles imply the places where the $r$-qubit gate acts on.
Each $r$-qubit gate is diagonal in the computational basis with random phases, i.e., ${\rm diag} (e^{i \phi_1}, e^{i \phi_2}, \cdots, e^{i \phi_{2^r}})$
and $\phi_k \in [0,2 \pi)$.
Note that the number of $r$-qubit diagonal gates is given by $\binom{N}{r}$.
}
\label{Fig:PRC}
\end{figure}

We present our result that an $r$-qubit phase-random circuit achieves a diagonal-unitary $t$-design, where $r$ is determined by $t$.
An $r$-qubit phase-random circuit is an extension of a phase-random circuit~\cite{NM2012,NTM2012}.  An $r$-qubit phase-random circuit for an $N$-qubit system is a quantum circuit consisting of $r$-qubit diagonal gates in the computational basis with random phases applied on all combinations of $r$ qubits out of $N$ qubits (see also Fig.~\ref{Fig:PRC}).  Note that each $r$-qubit gate {\it cannot} be decomposed into a sequence of $s$-qubit {\it diagonal} gates ($s<r$) since the phases of $r$-qubit diagonal gates should be chosen independently and randomly, which cannot be achieved by randomizing the phases of gates acting only on $s<r$ qubits.

Our first result on an implementation of diagonal-unitary $t$-designs is given in the following theorem.

\begin{Theorem} \label{Thm:NecSuf_r}
An $r$-qubit phase-random circuit is a diagonal-unitary $t$-design if and only if
$r \geq \lfloor \log_2 t \rfloor +1 $ for $t \leq 2^{N}-1$, and $r=N$ for $t \geq 2^N$.
\end{Theorem}

This result implies that we cannot achieve a diagonal-unitary $t$-design for $t \geq 4$ if we use only two- and one-qubit diagonal gates.  

Our second result is that each $r$-qubit gate for implementing a diagonal-unitary $t$-design can be replaced by
a sequence of multi-qubit controlled-phase-type gates that act on $s$ qubits ($s \leq r$), where
a multi-qubit controlled-phase-type gate is a unitary operation represented by ${\rm diag} ( 1, 1, \cdots, 1, e^{i \alpha})$ in the computational basis.
We also prove that a phase of a multi-qubit controlled-phase gate acting on $s$ qubits can be randomly selected from a $(\lfloor t/2^{s-1} \rfloor + 1)$-valued
discrete set of phases.   Thus, the smaller number of discrete phases is required for the gates acting on the larger number of qubits.
This result also shows that a phase-random circuit achieves a diagonal-unitary $t$-design with a finite number of elements.

These results enable us to analyze implementations of diagonal-unitary $t$-designs by using two- and one-qubit {\it non-diagonal} gates.
An explicit construction of a multi-qubit controlled-phase-type gate acting on $r$ qubits is known and it requires $O(r^2)$ two-qubit gates~\cite{BBCDMSSSW1995},
although it is unlikely to be optimal.
By decomposing the multi-qubit controlled-phase gates in an $r$-qubit phase-random circuit, we can show that
a diagonal-unitary $t$-design for an $N$-qubit system is obtained after applying $M$ two-qubit non-diagonal gates, where
\begin{equation*}
M= \binom{N}{r} \sum_{s=1}^{r-1} O(s^2) \binom{r}{s},
\end{equation*}
and $r$ is given in Theorem~\ref{Thm:NecSuf_r}.
For a constant $t$, $M \sim O(N^{\log_2 t})$ and the construction is efficient.
For larger $t$ such as $t={\rm poly}(N)$, this provides a sub-exponential implementation of a diagonal-unitary $t$-design.

\section{Proofs} \label{S:DandP}

We present proofs of all statements presented in Sec.~\ref{S:MR}.
In Subsec.~\ref{SS:Protocol}, 
we show the proof of Proposition~\ref{Prop:Protocol1}.
We present implementations of a diagonal-unitary $t$-design by an $r$-qubit phase-random circuit in Subsec.~\ref{SS:CI} and 
show a decomposition of each $r$-qubit gate into multi-qubit controlled-phase gates in Subsec.~\ref{SS:Decomp}.

Before presenting the proofs, we introduce our notation.
We denote $t$ $N$-bit sequences by $\mathbf{n}:=(\vec{n}^{(1)}, \cdots, \vec{n}^{(t)})$, where $\vec{n}^{(k)}$ is an $N$-bit sequence for $k=1,\cdots, t$,
and a set of all $\mathbf{n}$ by $\mathcal{D}$, i.e., $\mathcal{D}=\{ (\vec{n}^{(1)}, \cdots, \vec{n}^{(t)})
| \vec{n}^{(k)} \in \{0,1\}^{\times N} \text{ for } k=1,\cdots,t \}$.
Let $\mathcal{P}_t$ be a permutation group of order $t$.
We introduce an equivalent relation in $\mathcal{D}$ by $\mathcal{P}_t$; $\mathbf{n} \sim \mathbf{m}$ if and only if 
there exists $\sigma \in \mathcal{P}_t$ such that $\vec{n}^{(i)} = \vec{m}^{(\sigma(i))}$ for all $i \in \{1,\cdots, t\}$.
We denote by $\mathcal{D}/\mathcal{P}_t$ a quotient set of $\mathcal{D}$ by the equivalent relation.
For simplicity, we choose a representative of each equivalent class by $\mathbf{n}$ that satisfies
$\vec{n}^{(i)} \leq \vec{n}^{(i+1)}$ for every $i \in \{1,\cdots, t-1 \}$.
Note that the inequality is taken in binary, namely,
$\vec{n}^{(i)} \leq \vec{n}^{(j)}$ if and only if $\sum_{k=1}^N \vec{n}^{(i)}_k 2^{N-k} \leq \sum_{k=1}^N \vec{n}^{(j)}_k 2^{N-k}$, where 
$\vec{n}^{(i)}_k \in \{0,1\}$ is the $k$th bit of $\vec{n}^{(i)}$.
We finally introduce a canonical map $\pi$ from $\mathcal{D}$ to $\mathcal{D}/\mathcal{P}_t$,
and define $\pi^{-1}(\mathbf{n}') = \{ \mathbf{n} \in \mathcal{D} | \pi(\mathbf{n})=\mathbf{n}' \}$ for $\mathbf{n}' \in \mathcal{D}/\mathcal{P}_t$.
Using this notation, we define for $\mathbf{n}' \in \mathcal{D}/\mathcal{P}_t$,
\begin{equation*}
\ket{\pi^{-1}(\mathbf{n}')} := \frac{1}{\sqrt{ | \pi^{-1}(\mathbf{n}') |}} 
\sum_{\mathbf{m} \in \pi^{-1}(\mathbf{n}')} \ket{\mathbf{m}},
\end{equation*}
where $| \pi^{-1}(\mathbf{n}') |$ is the number of elements in $\pi^{-1}(\mathbf{n}') $.

\subsection{Generating an approximate state $t$-design by a diagonal-unitary $t$-design} \label{SS:Protocol}

We show Propositions~\ref{Prop:Protocol1} given in Sec.~\ref{S:MR}.

\begin{Proof}
The expectation of $\ketbra{\phi}{\phi}^{\otimes t}$ over random states is given by
\begin{equation*}
\mathbb{E}_{\ket{\phi} \in \Upsilon_{\rm Haar}} [\ketbra{\phi}{\phi}^{\otimes t}] = \frac{1}{\binom{t+d-1}{t}}\sum_{\mathbf{n}' \in \mathcal{D}/\mathcal{P}_t}  
\ketbra{\pi^{-1}(\mathbf{n}')}{\pi^{-1}(\mathbf{n}')}.
\end{equation*}
This is obtained by simply applying Schur's lemma~\cite{SW1986}.
For a $t$-design of phase-random states $\Upsilon_{\rm phase}$ obtained by applying a diagonal-unitary $t$-design on $\ket{+}^{\otimes N}$,
the expectation is given by
\begin{equation*}
\mathbb{E}_{\ket{\phi} \in \Upsilon_{\rm phase}} [\ketbra{\phi}{\phi}^{\otimes t}] = \frac{1}{d^t}\sum_{\mathbf{n}' \in \mathcal{D}/\mathcal{P}_t}  
| \pi^{-1}(\mathbf{n}') | \ketbra{\pi^{-1}(\mathbf{n}')}{\pi^{-1}(\mathbf{n}')}.
\end{equation*}

The difference $\eta(N,t)$ between the expectations is given by
\begin{align*}
\eta(N,t) &=
|\! | \mathbb{E}_{\ket{\phi} \in \Upsilon_{\rm Haar}}[\ketbra{\phi}{\phi}^{\otimes t}]-  \mathbb{E}_{\ket{\phi} \in \Upsilon_{\rm phase}}[\ketbra{\phi}{\phi}^{\otimes t}] |\! |_1 \\
&=
\sum_{\mathbf{n}' \in \mathcal{D}/\mathcal{P}_t}
\biggl| \frac{1}{\binom{t+d-1}{t}} - \frac{1}{d^{t}}  | \pi^{-1}(\mathbf{n}') | \biggr|.
\end{align*}
We expand $\binom{t+d-1}{t}$ by $\frac{1}{t!}\sum_{k=1}^t \alpha_k d^k$, where $\alpha_t=1$ and $\alpha_{t-1}=t(t-1)/2$, and obtain
\begin{equation}
\eta(N,t)=\biggl( \sum_{k=1}^{t} \alpha_k d^{k-t} \biggr)^{-1}
\sum_{\mathbf{n}' \in \mathcal{D}/\mathcal{P}_t}
\biggl|
t! \frac{1}{d^t} - | \pi^{-1}(\mathbf{n}') | \sum_{k=1}^{t} \alpha_{k}  \frac{1}{d^{2t-k}} 
\biggr|. \label{Eq:;lkjfsda}
\end{equation}
We explicitly calculate this up to the order $1/d$.
For $d \gg 1$ and a constant $t$, $( \sum_{k=1}^{t} \alpha_k d^{k-t} )^{-1}$ is given by 
\begin{equation}
\biggl( \sum_{k=1}^{t} \alpha_k d^{k-t} \biggr)^{-1} = 1- \frac{t(t-1)}{2} \frac{1}{d} + O(\frac{1}{d^2}). \label{Eq;oi}
\end{equation}
The other term in Eq.~\eqref{Eq:;lkjfsda}, $| \pi^{-1} (\mathbf{n}') |$, depends only on the number of the same $N$-bit sequences in $\mathbf{n}'$.
For $\mathbf{n}'$ such that $\vec{n}^{(i)} \neq \vec{n}^{(j)}$ for $i\neq j$, $| \pi^{-1}(\mathbf{n}') |=t!$ and the number of such $\mathbf{n}'$ is $\binom{d}{t}$.
For $\mathbf{n}'$ such that there exists only one pair $(i,j)$, where $i\neq j$, satisfying $\vec{n}^{(i)} = \vec{n}^{(j)}$, 
$| \pi^{-1}(\mathbf{n}') |=t!/2$ and the number of such $\mathbf{n}'$ is $(t-1)\binom{d}{t-1}$.
In other cases, the number of each type of $\mathbf{n}'$ is at most $d^{t-2}$.
Since the inside of the summation $\sum_{\mathbf{n}' \in \mathcal{D}/\mathcal{P}_t}$ in Eq.~\eqref{Eq:;lkjfsda} is at most $O(1/d^{t})$, we obtain that 
\begin{align}
&\sum_{\mathbf{n}' \in \mathcal{D}/\mathcal{P}_t}
\biggl|
t! \frac{1}{d^t} - | \pi^{-1}(\mathbf{n}') | \sum_{k=1}^{t} \alpha_{k}  \frac{1}{d^{2t-k}} 
\biggr| \notag \\ 
&=\binom{d}{t}\biggl( t! \frac{t(t-1)}{2} \frac{1}{d^{t+1}} + O(\frac{1}{d^{t+2}})  \biggr) +
(t-1)\binom{d}{t-1}\biggl( \frac{t!}{2} \frac{1}{d^{t}} + O(\frac{1}{d^{t+1}})  \biggr) + O(\frac{1}{d^{2}}) \notag \\
&=\frac{t(t-1)}{d} + O(\frac{1}{d^2}), \label{Eq:2;l43m}
\end{align}
where we have used relations such as $\binom{d}{t} = ( d^t + O(d^{t-1}) )/t!$ for a constant $t$.
From Eqs.~\eqref{Eq;oi} and~\eqref{Eq:2;l43m}, we obtain
\begin{equation*}
\eta(N,t)=\frac{t(t-1)}{d} + O(\frac{1}{d^2}),
\end{equation*}
\begin{flushright}$\blacksquare$\end{flushright}
\end{Proof}

\subsection{An $r$-qubit phase-random circuit achieves a diagonal-unitary $t$-design} \label{SS:CI}

We prove Theorem~\ref{Thm:NecSuf_r} by restating it in a different way.
Since the statement is obvious for $r=N$, we consider only $r \leq N-1$ in the following.

Let $S$ be $\mathbb{E}_{U \in \mathcal{U}_{\rm diag}} [ U^{\otimes t} \otimes (U^{\dagger} ) ^{\otimes t}]$ and introduce expansion coefficients $S_{\mathbf{n} \mathbf{m}}$ in the computational basis defined by
\begin{equation*}
S=\sum_{\mathbf{n}, \mathbf{m} \in \mathcal{D}} S_{\mathbf{n} \mathbf{m}} \ketbra{\mathbf{n}}{\mathbf{n}} \otimes \ketbra{\mathbf{m}}{\mathbf{m}}.
\end{equation*}
Note that $S$ is diagonal in the computational basis since $S$ is an expectation of $U^{\otimes t} \otimes (U^{\dagger} ) ^{\otimes t}$ for $U \in \mathcal{U}_{\rm diag}$, and $U \in \mathcal{U}_{\rm diag}$ is diagonal in the computational basis.
A simple calculation leads to 
\begin{equation}
S_{\mathbf{n} \mathbf{m}}=
\begin{cases}
1 & \text{when  } \pi(\mathbf{n}) = \pi(\mathbf{m}), \\
0 & \text{when  } \pi(\mathbf{n}) \neq \pi(\mathbf{m}).
\end{cases} \label{Eq:Av}
\end{equation}
Our goal is to derive the value of $r$ for which the $r$-qubit phase-random circuit achieves the coefficients $S_{\mathbf{n} \mathbf{m}}$.

We denote by $I_s$ a subset of $\{1,\cdots , N\}$ with $s$ elements.
For a given $I_s=\{i_1,\cdots,i_s\}$,
we denote $s$-bit subsequences in $\vec{n}^{(k)}$ and $\vec{m}^{(k)}$ at $I_s$ by $\vec{n}^{(k)}_{I_s}:=n^{(k)}_{i_1} \cdots n^{(k)}_{i_s}$
and $\vec{m}^{(k)}_{I_s}:= m^{(k)}_{i_1} \cdots m^{(k)}_{i_s}$, respectively,
and 
$(\vec{n}^{(1)}_{I_s}, \cdots,  \vec{n}^{(t)}_{I_s})$ and $(\vec{m}^{(1)}_{I_s}, \cdots,  \vec{m}^{(t)}_{I_s})$
by $\mathbf{n}_{I_s}$ and $\mathbf{m}_{I_s}$, respectively.
We generalize a canonical map $\pi$ to the one mapping $t$ $s$-bit sequences $\mathcal{D}_s$ to a quotient set $\mathcal{D}_s/\mathcal{P}_t$  by 
the permutation group $\mathcal{P}_t$. In this notation, $\mathcal{D}_N= \mathcal{D}$.
We call the number of $1$ in a bit sequence {\it weight} of the sequence.
By using these expressions, we first prove the following lemma.

\begin{Lemma} \label{Lemma:rgfda}
Let $\mathbf{n}, \mathbf{m} \in \mathcal{D}_s$ be such that
$\pi(\mathbf{n}) \neq \pi(\mathbf{m})$
and $\pi(\mathbf{n}_{I_{s-1}}) = \pi(\mathbf{m}_{I_{s-1}})$ for any $I_{s-1} \subset \{1,\cdots, s\}$.
Denote by $G_{\mathbf{n}} (\vec{n})$ the number of $\vec{n}$ in $\mathbf{n}$. 
Then, for any $s$-bit sequence $\vec{n}$,
\begin{equation}
| G_{\mathbf{n}} (\vec{n}) - G_{\mathbf{m}} (\vec{n}) | = g,  \label{Eq:p34t}
\end{equation}
where $g$ is a constant positive integer.
Moreover, $t \geq 2^{s-1} g$.
\end{Lemma}

\begin{Proof}
If $\exists q, q' \in \{1,\cdots, t\}$ such that $\vec{n}^{(q)}= \vec{m}^{(q')}$,
we remove them from $\mathbf{n}$ and $\mathbf{m}$.
As a result, we obtain $\tilde{\mathbf{n}}$ and $\tilde{\mathbf{m}}$,
which are composed of $t'$ $s$-bit sequences for $t' \leq t$.
Note that $\tilde{\mathbf{n}}$ and $\tilde{\mathbf{m}}$ 
still satisfy
$\pi(\tilde{\mathbf{n}}_{I_{s-1}}) = \pi(\tilde{\mathbf{m}}_{I_{s-1}})$ for any $I_{s-1} \subset \{1,\cdots, s\}$.

Without loss of generality, we assume that 
the $s$-bit sequence with the most occurrence in $\tilde{\mathbf{n}}$ is $00 \cdots 0$, and let $g$ be the number of the occurrence.
Since $\pi(\tilde{\mathbf{n}}_{I_{s-1}}) = \pi(\tilde{\mathbf{m}}_{I_{s-1}})$ for any $I_{s-1} \subset \{1, \cdots, s\}$
and all $s$-bit sequences in $\tilde{\mathbf{n}}$ differ from those in $\tilde{\mathbf{m}}$,
all sequences with weight one should be contained in $\tilde{\mathbf{m}}$.
Moreover, the number of each sequence with weight one in $\tilde{\mathbf{m}}$ is $g$ since 
the number of $00\cdots 0$ in $\tilde{\mathbf{n}}$ is $g$.
This in turn implies that all sequences with weight two should be contained in $\tilde{\mathbf{n}}$.
Similarly, the number of each of such sequences should be $g$.
By repeating this, it follows that all sequences with zero or even weight are contained in $\tilde{\mathbf{n}}$,
and those with odd weight are in $\tilde{\mathbf{m}}$.
In addition, the number of each sequence is $g$.
Thus, we obtain for any $s$-bit sequence $\vec{n}$,
\begin{equation*}
| G_{\tilde{\mathbf{n}}} (\vec{n}) - G_{\tilde{\mathbf{m}}} (\vec{n}) | = g, 
\end{equation*}
and $t' = 2^{s-1} g$.
By construction, $| G_{\tilde{\mathbf{n}}} (\vec{n}) - G_{\tilde{\mathbf{m}}} (\vec{n}) | = | G_{\mathbf{n}} (\vec{n}) - G_{\mathbf{m}} (\vec{n}) |$,
so that we obtain Eq.~\eqref{Eq:p34t}. Moreover, as $t \geq t'$, it follows that $t \geq 2^{s-1} g$.
\begin{flushright}$\blacksquare$\end{flushright}
\end{Proof}

By using Lemma~\ref{Lemma:rgfda}, we show the following Proposition.

\begin{Proposition} \label{Prop:asbarag}
For $r \leq N-1$, the following are equivalent;
\begin{enumerate}
\item[(A)] An $r$-qubit phase-random circuit achieves an exact diagonal-unitary $t$-design, 
\item[(B)] For any $\mathbf{n}$, $\mathbf{m} \in \mathcal{D}$,
if $\pi(\mathbf{n}_{I_r}) =\pi(\mathbf{m}_{I_r})$ for any $I_r \subset \{1,\cdots, N \}$,
then $\pi(\mathbf{n}) =  \pi(\mathbf{m})$,
\item[(C)] $r > \log_2t$.
\end{enumerate}
\end{Proposition}
From the equivalence of (A) and (C), we obtain Theorem~\ref{Thm:NecSuf_r}.

\begin{Proof}
We first show the equivalence of (A) and (B), and then that of (B) and (C).
The unitary matrix corresponding to an $r$-qubit phase-random circuit is given by 
\begin{equation*}
W_{\phi} = \prod_{I_r  \subset \{1,\cdots, N\} } W_{I_r}, 
\end{equation*}
where $W_{I_r} := {\rm diag}_{I_r} (e^{i \phi_1}, \cdots, e^{i \phi_{2^r}} ) \otimes \mathbb{I}_{\{1,\cdots,N\} \setminus I_r}$ is a diagonal unitary matrix with random phases $\{\phi_1, \cdots, \phi_{2^r} \}$ acting non-trivially on the qubits at sites $I_r$.
The matrix $\mathbb{I}_{I}$ represents the identity matrix acting on qubits at sites $I \subset \{1,\cdots, N\}$.
Since the random phases are independently chosen for each $W_{I_r}$, the expectation of $W_{\phi}^{\otimes t} \otimes (W^{\dagger}_{\phi})^{\otimes t} $ over all random phases is given by
\begin{align*}
\mathbb{E}[W_{\phi}^{\otimes t} \otimes (W^{\dagger}_{\phi})^{\otimes t}] &=  
\prod_{I_r \subset \{ 1, \cdots, N \}} \mathbb{E}[ W_{I_r}^{\otimes t} \otimes (W_{I_r}^{\dagger})^{\otimes t}] \\
&= \prod_{I_r \subset \{1, \cdots, N \}} \sum_{\mathbf{n}, \mathbf{m} \in \mathcal{D}} W^{\mathbf{n} \mathbf{m}}_{I_r} \ketbra{\mathbf{n}}{\mathbf{n}} \otimes \ketbra{\mathbf{m}}{\mathbf{m}},
\end{align*}
where 
\begin{equation}
W^{\mathbf{n} \mathbf{m}}_{I_r}
=
\begin{cases}
1 & \text{when  } \pi(\mathbf{n}_{I_r}) = \pi(\mathbf{m}_{I_r}), \\
0 & \text{when  } \pi(\mathbf{n}_{I_r}) \neq \pi(\mathbf{m}_{I_r}).
\end{cases} \label{Eq:ProdAv}
\end{equation}
By comparing Eq.~\eqref{Eq:ProdAv} with Eq.~\eqref{Eq:Av}, we obtain the equivalence of (A) and (B).

Next, we show that (C) implies (B) by showing its contraposition, namely,
if there exists $\mathbf{n}$, $\mathbf{m} \in \mathcal{D}$ that satisfies $\pi(\mathbf{n}) \neq \pi(\mathbf{m})$ but $\forall I_r \subset  \{1,\cdots, N \}$,
$\pi(\mathbf{n}_{I_r}) = \pi(\mathbf{m}_{I_r})$, then $t \geq 2^r$.
By assumption, there exists $r' \geq r$ and $I_{r'+1}$ such that $\pi(\mathbf{n}_{I_{r'+1}}) \neq \pi(\mathbf{m}_{I_{r'+1}})$ and $\pi(\mathbf{n}_{I_{r'}}) = \pi(\mathbf{m}_{I_{r'}})$ for any $I_{r'} \subset I_{r'+1}$.
It follows from Lemma~\ref{Lemma:rgfda} that $t \geq 2^{r'} g$, where $g \geq 1$.
As $r' \geq r$, we obtain $t \geq 2^r$.

Finally, we show that (B) implies (C). This is also obtained by showing its contraposition.
Consider 
$\mathbf{n}$ and $\mathbf{m} \in \mathcal{D}$ such that, for a fixed $I_{r+1}$, $\mathbf{n}_{I_{r+1}}$ and $\mathbf{m}_{I_{r+1}}$ contain all $(r+1)$-bit sequences with even or zero weight and those with odd weight, respectively, and $\mathbf{n}_{\{1,\cdots, N\} \setminus I_{r+1}} = \mathbf{m}_{\{1,\cdots, N\} \setminus I_{r+1}}$.
Such $\mathbf{n}$ and $\mathbf{m}$ exist if $t \geq 2^r$.
It is obvious that $\pi(\mathbf{n}) \neq \pi(\mathbf{m})$.
However, it is easy to see that $\pi(\mathbf{n}_{I_r}) = \pi(\mathbf{m}_{I_r})$ for any $I_r \subset \{1,\cdots, N \}$.
This shows the contraposition of the statement that (B) implies (C), and concludes the proof.
\begin{flushright}$\blacksquare$\end{flushright}
\end{Proof}

\subsection{Decomposition of $r$-qubit gates into the controlled-phase-type gates} \label{SS:Decomp}
We show how to decompose each $r$-qubit diagonal gate in an $r$-qubit phase-random circuit into a sequence of multi-qubit controlled-phase-type gates with discrete random phases.
More precisely, we prove the following.

\begin{Proposition}
To implement a diagonal-unitary $t$-design by an $r$-qubit phase-random circuit, every gate ${\rm diag}_{I_r} ( 1, e^{i \phi_1}, \cdots, e^{i \phi_{2^r-1}}) \otimes \mathbb{I}_{\{1,\cdots, N\} \setminus I_r}$ non-trivially acting on $r$ qubits at $I_r$
with random phases $\phi_k \in [0, 2 \pi)$ for $k=1, \cdots, 2^r-1$ can be replaced by
\begin{equation*}
D_{I_r} = \prod_{I_s \subset I_r, 1\leq s \leq r} {\rm diag}_{I_s} ( 1, 1, \cdots, 1, e^{i \alpha_{I_s}}) \otimes \mathbb{I}_{\{1,\cdots, N\} \setminus I_s}, 
\end{equation*}
where $\alpha_{I_s}$ are randomly and independently chosen from
\begin{equation}
\biggl\{ \frac{2 \pi k}{\lfloor t/2^{s-1}   \rfloor + 1} \biggr\}_{k=0, 1, \cdots, \lfloor t/2^{s-1}   \rfloor }. \label{Eq:DrPhase}
\end{equation}
\end{Proposition}

\begin{Proof}
We consider an $r$-qubit diagonal gate acting on qubits at $I_r$. Since 
it is sufficient to consider a nontrivial part of the matrix,
we investigate a diagonal matrix given by 
\begin{equation*}
W_{I_r} = \sum_{\vec{n}_{I_r}} e^{i \phi_{\vec{n}_{I_r}}} \ketbra{\vec{n}_{I_r}}{\vec{n}_{I_r}}.
\end{equation*}
Its tensor product $W_{_{I_r}}^{ \otimes t} \otimes (W_{_{I_r}}^{\dagger})^{ \otimes t}$ is given by
\begin{equation*}
W_{I_r}^{ \otimes t} \otimes (W_{I_r}^{\dagger})^{ \otimes t}
=
\sum_{\mathbf{n}_{I_r}, \mathbf{m}_{I_r} \in \mathcal{D}_{r}} e^{i \phi_{\mathbf{n}_{I_r} \mathbf{m}_{I_r}}} \ketbra{\mathbf{n}_{I_r}}{\mathbf{n}_{I_r}} \otimes \ketbra{\mathbf{m}_{I_r}}{\mathbf{m}_{I_r}},
\end{equation*}
where  
$\phi_{\mathbf{n}_{I_r} \mathbf{m}_{I_r}} = \sum_{k=1}^t (\phi_{\vec{n}^{(k)}_{I_r}} - \phi_{\vec{m}^{(k)}_{I_r}})$.
If every phase is randomly chosen from $[0, 2 \pi)$, 
\begin{equation}
\mathbb{E}[e^{i \phi_{\mathbf{n}_{I_r} \mathbf{m}_{I_r}}}]=
\begin{cases}
1 & \text{ when }  \pi(\mathbf{n}_{I_r}) = \pi(\mathbf{m}_{I_r}), \\
0 & \text{ when }  \pi(\mathbf{n}_{I_r}) \neq \pi(\mathbf{m}_{I_r}).
\end{cases} \label{Eq:13r}
\end{equation}

We investigate the coefficient of $\ketbra{\mathbf{n}_{I_r}}{\mathbf{n}_{I_r}} \otimes \ketbra{\mathbf{m}_{I_r}}{\mathbf{m}_{I_r}}$ in $D_{I_r}^{\otimes t}\otimes (D_{I_r}^{\dagger})^{\otimes t}$.
Our goal is to show that the choice of the phases defined in Eq.~\eqref{Eq:DrPhase} achieves the same average as Eq.~\eqref{Eq:13r}.
For this purpose, it is sufficient to prove the following two properties:
(a) For any $I_1 \subset I_r$, the average over $\alpha_{I_1}$ makes the coefficient of $\ketbra{\mathbf{n}_{I_1}}{\mathbf{n}_{I_1}} \otimes \ketbra{\mathbf{m}_{I_1}}{\mathbf{m}_{I_1}}$ vanish if $\pi(\mathbf{n}_{I_1}) \neq \pi(\mathbf{m}_{I_1})$. 
(b) For any $I_s \subset I_r$ with any $1<s \leq r$, the average over $\alpha_{I_s}$ makes the coefficients of $\ketbra{\mathbf{n}_{I_s}}{\mathbf{n}_{I_s}} \otimes \ketbra{\mathbf{m}_{I_s}}{\mathbf{m}_{I_s}}$ vanish if $\pi(\mathbf{n}_{I_s})\neq \pi(\mathbf{m}_{I_s})$ and 
  $\pi(\mathbf{n}_{I_{s-1}})= \pi(\mathbf{m}_{I_{s-1}})$ for all $I_{s-1}\subset I_s$.

The property (a) holds since
the coefficients of $\ketbra{\mathbf{n}_{I_1}}{\mathbf{n}_{I_1}} \otimes \ketbra{\mathbf{m}_{I_1}}{\mathbf{m}_{I_1}}$ in 
$D_{I_r}^{\otimes t}\otimes (D_{I_r}^{\dagger})^{\otimes t}$
includes the factor $e^{i t' \alpha_{I_1}}$ with $1 \leq t' \leq t$, which vanishes after taking the average of $\alpha_{I_1}$ over 
$\{ \frac{2 \pi k}{ t + 1}\}_{k=0, 1, \cdots, t }$.
The property (b) is obtained from Lemma~\ref{Lemma:rgfda}.
When $\mathbf{n}_{I_s}$ and $\mathbf{m}_{I_s}$ satisfy $\pi(\mathbf{n}_{I_s})\neq \pi(\mathbf{m}_{I_s})$ and 
  $\pi(\mathbf{n}_{I_{s-1}})= \pi(\mathbf{m}_{I_{s-1}})$ for all $I_{s-1}\subset I_s$,
it follows from Lemma~\ref{Lemma:rgfda} that, for an $s$-bit sequence $\vec{1}_s :=11\cdots 1$,
\begin{equation*}
| G_{\mathbf{n}_{I_s}} (\vec{1}_{s}) - G_{\mathbf{m}_{I_s}} (\vec{1}_{s}) | = g.
\end{equation*}
This implies that the coefficient of $\ketbra{\mathbf{n}_{I_s}}{\mathbf{n}_{I_s}} \otimes \ketbra{\mathbf{m}_{I_s}}{\mathbf{m}_{I_s}}$ for such $\mathbf{n}_{I_s}$ and $\mathbf{m}_{I_s}$
contains $e^{i g \alpha_{I_s}}$ or $e^{-i g \alpha_{I_s}}$.
We also obtain from Lemma~\ref{Lemma:rgfda} that $t \geq 2^{s-1} g$, namely, $g \leq \lfloor t/2^{s-1}  \rfloor$,
where the equality holds for $\mathbf{n}_{I_s}$ and $\mathbf{m}_{I_s}$ that do not share the same $s$-bit sequences.
Thus, for these terms to be zero by taking the average over $\alpha_{I_s}$, it is sufficient to take $\alpha_{I_s}$ randomly and independently from
\begin{equation*}
\biggl\{ \frac{2\pi k}{ \lfloor t/2^{s-1}   \rfloor +1} \biggr\}_{k=0,\cdots, \lfloor t/2^{s-1}   \rfloor}.
\end{equation*}
\begin{flushright}$\blacksquare$\end{flushright}
\end{Proof}



\section{Summary and concluding remarks} \label{S:Conc}

We investigated protocols of generating a state $t$-design in an $N$-qubit system by using 
a diagonal-unitary $t$-design in the computational basis applied on a fixed separable state.
We have first shown that a $O(1/2^N)$-approximate state $t$-design is generated by simply applying the diagonal-unitary $t$-design.
We have then investigated a way of improving the degree of approximation
by exploiting a local random circuit in addition to a diagonal-unitary $t$-design, 
which seems to result in a faster convergence than the protocol using only a local random circuit.
We have also investigated quantum circuit implementations of a diagonal-unitary $t$-design, and have shown that
an $r$-qubit phase-random circuit, where $r \geq \lfloor \log_2 t \rfloor +1$ for $t \leq 2^N-1$ and $r=N$ for $t \geq 2^N$,
generates a diagonal-unitary $t$-design. The number of $r$-qubit gates in the circuit is given by $\binom{N}{r}$.
Each $r$-qubit diagonal gate has been shown to be decomposable into a sequence of $s$-qubit multi-qubit controlled-phase gates ($s \leq r$) with 
$(\lfloor t/2^{s-1} \rfloor +1)$-valued
discrete random phases.

We make remarks on possible future directions.
First, numerical analysis of the method applying a local random circuit in addition to a diagonal-unitary $t$-design
is less conclusive, so that further numerical or analytical investigations are required.
For an analytical investigation, it is sufficient to check how the coefficients of $\ketbra{\pi^{-1}(\mathbf{n})}{\pi^{-1}(\mathbf{n})}$
are changed by a local random circuit.
Although we studied a protocol of generating a {\it state} $t$-design from a fixed initial state in this paper,
it is interesting to investigate if our protocol also achieves a {\it unitary} $t$-design more efficiently.

Another direction is to deepen the analysis of a quantum circuit implementation of a diagonal-unitary $t$-design by
using {\it non-diagonal} gates.
We have provided an implementation of an exact diagonal-unitary $t$-design by using $O(N^{\log t})$ non-diagonal two-qubit gates for a constant $t$.
However, the scaling is worse for a large $t$ than that of a unitary $t$-design implemented by a local random circuit,
which requires $O(N t^4 (N + \log1/\epsilon))$ non-diagonal two-qubit gates~\cite{BHH2012}.
Since our implementation is probably not optimal, it is interesting to see a lower bound of the length of the circuit to implement a diagonal-unitary $t$-design.

It will be also interesting to investigate in which cases a unitary $t$-design used in a quantum protocol or task can be subsituted by a diagonal-unitary $t$-design.
We have shown in this paper that generation of a state $t$-design is one of the cases.
Random unitary matrices have been also exploited for decoupling two systems~\cite{DBWR2010,BF2013}.
Since decoupling has many applications in quantum information processing, it may be interesting and useful
to investigate if diagonal-unitary $t$-designs are capable to achieve an exact or approximate decoupling.

\section{Acknowledgment}
This work is supported by Project for Developing Innovation Systems of the Ministry of Education, Culture, Sports, Science and Technology (MEXT), Japan. 
Y.~N. and M.~M acknowledge support from JSPS by KAKENHI, Grant No. 222812 and 23540463, respectively.
Y. N. also acknowledges JSPS Postdoctoral Fellowships for Research Abroad.
M.~K is supported by the Funding Program for World-Leading
Innovative R\&D on Science and Technology (FIRST).

\appendix

\section{Generating a state $t$-design by a diagonal-unitary $t$-design and a probabilistic procedure} \label{Ap:classical}

We consider a protocol of generating a state $t$-design by using a diagonal-unitary $t$-design and a classical probabilistic procedure,
and show that the improvement of the degree of approximation is limited to be $O(d^{1-t})$.

We generalize a protocol introduced in Ref.~\cite{NM2012} as follows:

\begin{enumerate}
\item With probability $p$, apply a diagonal-unitary $t$-design on $\ket{+}^{\otimes N}$.
\item With probability $1-p$, choose a random $N$-bit sequence $\vec{n}$ and generate $\ket{\vec{n}}$.
\end{enumerate}
We denote by $\Upsilon(p)$ the resulting ensemble and show that
\begin{equation*}
\min_{p \in [0,1]} 
\biggl|\!\biggl|
\mathbb{E}_{\ket{\psi} \in \Upsilon_{\rm Haar}} [ \ketbra{\psi}{\psi}^{\otimes t} ] 
- 
\mathbb{E}_{\ket{\psi} \in \Upsilon(p)} [\ketbra{\psi}{\psi}^{\otimes t}] \biggr|\! \biggr|_{1} 
= \eta(N,t) - O(\frac{1}{d^{t-1}}),
\end{equation*}
where the minimum is given by 
\begin{equation*}
p = \frac{1 - d/\binom{t+d-1}{t}}{1- d^{1-t}}.
\end{equation*}

Following the calculations in Subsec.~\ref{SS:Protocol}, the difference between 
the expectation over random states and that over $\Upsilon(p)$ is given by
\begin{align*}
D(p)&:=|\! | \mathbb{E}_{\ket{\psi} \in \Upsilon_{\rm Haar}}[\ketbra{\phi}{\phi}^{\otimes t}]-  \mathbb{E}_{\ket{\phi} \in \Upsilon(p)}[\ketbra{\phi}{\phi}^{\otimes t}] |\! |_1 \\
&=
d \biggl| \frac{1}{\binom{t+d-1}{t}} - (\frac{1-p}{d} + \frac{p}{d^t}) \biggr| + 
\sum_{\mathbf{n}' \in \mathcal{D'}/\mathcal{P}_t}
\biggl|\frac{1}{\binom{t+d-1}{t}} - \frac{p}{d^{t}}    | \pi^{-1}(\mathbf{n}') | \biggr|,
\end{align*}
where $\mathcal{D}'=\mathcal{D} \setminus \{(\vec{n},\vec{n},\cdots,\vec{n}) | \vec{n} \in \{0,1 \}^{\times N} \}$.
In this notation, $D(1)=\eta(N,t)$.

Since $D(p)$ is a linear function of $p$, it is sufficient to investigate the coefficient of $p$.
It is straightforward to observe the following:
\begin{equation*}
\frac{1}{\binom{t+d-1}{t}} - (\frac{1-p}{d} + \frac{p}{d^t}) 
\begin{cases}
<0 & \text{ when } p < \frac{1-d/\binom{t+d-1}{t}}{1-d^{1-t} }, \\
>0 & \text{ when } p > \frac{1-d/\binom{t+d-1}{t}}{1-d^{1-t} },
\end{cases}
\end{equation*}
for $\mathbf{n}'=(\vec{n}^{(1)}, \cdots, \vec{n}^{(t)})$, where $\vec{n}^{(i)} \neq \vec{n}^{(j)}$ for $i \neq j$,
\begin{equation*}
\frac{1}{\binom{t+d-1}{t}} - \frac{p}{d^{t}}    | \pi^{-1}(\mathbf{n}') |
\begin{cases}
> 0 & \text{ when } p < \frac{d^t}{\binom{t+d-1}{t} t!}, \\
< 0 & \text{ when } p > \frac{d^t}{\binom{t+d-1}{t} t!},
\end{cases}
\end{equation*}
and, for other $\mathbf{n}'$,
\begin{equation*}
\frac{1}{\binom{t+d-1}{t}} - \frac{p}{d^{t}}    | \pi^{-1}(\mathbf{n}') | >0.
\end{equation*}
By taking the absolute values into account, we obtain that 
the coefficient of $p$ is negative for $p < p_0$ and positive for $p>p_0$, where $p_0 = \frac{1-d/\binom{t+d-1}{t}}{1-d^{1-t} }$,
so that $D(p_0)$ is the minimum.
The order of $D(p_0)$ is easily estimated from a fact that 
$p_0 \sim 1 - (t!+1)d^{1-t}$.
Since $p$ is a probability of mixing a $t$-design of $\Upsilon(1)$ and a separable state $\{ \ketbra{\vec{n}}{\vec{n}} \}$,
$D(1) - D(p_0) = O(d^{1-t})$, resulting in $\min_p D(p) = \eta(N,t) - O(d^{1-t})$.

\end{document}